\documentclass[a4paper,12pt]{article}

\usepackage[cp1250]{inputenc} %Czech and Slovak symbols
\usepackage{authblk}%Formating of authors' information on title page
\usepackage{a4wide}
\usepackage{graphicx}
\usepackage{amsfonts}
\usepackage{amsmath}
\usepackage{amsthm}
\usepackage{amssymb}
\usepackage{booktabs}
\usepackage{multirow}
\usepackage{epstopdf}

\begin{document}

\title{What are the main drivers of the Bitcoin price? Evidence from wavelet coherence analysis}

\author[a,b]{Ladislav Kristoufek}
\affil[a]{Institute of Economic Studies, Faculty of Social Sciences, Charles University in Prague, Opletalova 26, 110 00, Prague, Czech Republic, EU}
\affil[b]{Institute of Information Theory and Automation, Academy of Sciences of the Czech Republic, Pod Vodarenskou Vezi 4, 182 08, Prague, Czech Republic, EU, tel.: +420266052243, e-mail: kristouf@utia.cas.cz}
\date{}
\maketitle

\begin{abstract}
\footnotesize
Bitcoin has emerged as a fascinating phenomenon of the financial markets. Without any central authority issuing the currency, it has been associated with controversy ever since its popularity and public interest reached high levels. Here, we contribute to the discussion by examining potential drivers of Bitcoin prices ranging from fundamental to speculative and technical sources as well as a potential influence of the Chinese market. The evolution of the relationships is examined in both time and frequency domains utilizing the continuous wavelets framework so that we comment on development of the interconnections in time but we can also distinguish between short-term and long-term connections.
\end{abstract}

\normalsize

\section{Introduction}

Bitcoin \cite{Nakamoto} is a potential alternative currency to the standard fiat currencies (US dollar, the Euro, Japanese Yen, etc.) with various advantages such as low or no fees, a controlled and known algorithm of the currency creation, and an informational transparency of all transactions. Its success has ignited an exposition of new alternative crypto-currencies, usually labelled as ``Altcoins'', with various motivations and aims. Bitcoin still remains a leading crypto-currency and with its massive capitalization of approximately 6 billion US dollars, it safely dominates the other crypto-currencies. Of course, where an upside is there usually is a downside as well. During its increasing popularity and public attention, the Bitcoin system has been accused and labelled as an environment for organized crime and money laundering, and it has been a target of repeated hackers attacks causing some major losses to the bitcoins owners \cite{Barratt2012}. However, it needs to be noted that none of these downsides is completely remote to the standard cash currencies either.

Even though Bitcoin has been frequently discussed on various financial blogs and even mainstream financial media, the research community is still mainly focused on the currency's technical, safety and legal issues \cite{Barratt2012,Jacobs2011,Barber2012,Clark2012,Reid2013,Velde2013} while discussion about the economic and financial aspects remains relatively sparse. Bornholdt \& Sneppen \cite{Bornholdt2014} construct a model with a voter-like dynamics and show that Bitcoin holds no special advantages above the other crypto-currencies and might as well be replaced by some alternative. Kondor \textit{et al.} \cite{Kondor2014} study the Bitcoin network in a standard complex networks framework and they show that the network characteristics of Bitcoin evolve in time and these are due to bitcoins being accepted as a mean of payment more frequently. Further, they show that the wealth in bitcoins is getting more accumulated in time and such accumulation is tightly related to the ability to attract new connections in the network. Garcia \textit{et al.} \cite{Garcia2014} study the Bitcoin bubbles using digital behavioral traces of investors by means of social media use, search queries and user base. They find positive feedback loops in the social media use and the user base. In our previous study \cite{Kristoufek2013}, we focus on a speculative part of the Bitcoin value measured by the search queries on Google and searched words on Wikipedia showing that both the bubble and bust cycles of the Bitcoin prices can be at least partially explained by the interest in the currency. Since then, Bitcoin has attracted even more attention as its exchange rate with the US dollar breached the \$1000 level (with a maximum of \$1242 per a bitcoin at the Mt. Gox market, creating an absurd potential profit of more than 9000\% for a buy-and-hold strategy in less than 11 months) in late November and early December 2013. After some following corrections, the value of Bitcoin has stabilized between \$900 and \$1000 per a bitcoin at a break of years 2013 and 2014. However, a huge strike to the Bitcoin credibility and reputation came with an insolvency of the Mt. Gox exchange, historically the most prominent of the Bitcoin markets, after which the Bitcoin price has started a slow stable decreasing trend with a rather low volatility and a bitcoin now (April 2014) trades around \$500.

Here, we tackle the price of the Bitcoin currency from a wider perspective. We focus on various possible sources of the price movements, ranging from fundamental to speculative and technical sources, and examine how the interconnections behave in time but also at different scales (frequencies). To do so, we utilize continuous wavelet analysis and specifically the wavelet coherence which is able to localize correlations between series, its evolution in time and across scales. Detailed description of wavelets framework used in the text is provided in the Methods section. It needs to be stressed that both time and frequency aspects are important for the Bitcoin price dynamics as the currency has undergone a wild evolution in recent years and it would be naive to believe that the driving forces of the prices have remained unchanged during its existence. In addition, the frequency domain viewpoint gives an opportunity to distinguish between short-term and long-term correlations. We show that indeed both time and frequency characteristics of the dynamics are worth the investigation and various interesting relationships are uncovered.

\section{Methods}

\subsection{Data}

\subsubsection*{Bitcoin price index}

Bitcoin price index (BPI) is an index of the exchange rate between the US dollar (USD) and Bitcoin (BTC). There are various criteria for specific exchanges to be included in BPI which are, as of now, met by three exchanges -- Bitfinex, Bitstamp and BTC-e. Historically, Mt. Gox exchange used to be part of the index as well but due to its closure, the criteria ceased to be fulfilled. BPI is available on a 1-min basis and it is formed as a simple average of the covered exchanges. The series are freely available at http://www.coindesk.com/price. Due to data availability, we analyze the relationships starting from 14 September 2011.

\subsubsection*{Blockchain}

Blockchain (http://www.blockchain.info) freely provides very detailed series about Bitcoin markets. On a daily basis, the following time series, which are used in our analysis, are reported:
\begin{itemize}
\item Total Bitcoins in circulation
\item Number of transactions excluding exchange transactions
\item Estimated output volume
\item Trade volume vs. transaction volume ratio
\item Hash rate
\item Difficulty
\end{itemize}

Total number of bitcoins in circulations is given by a known algorithm and asymptotically, it reaches 21 million bitcoins. The creation of new bitcoins is driven and regulated by difficulty which mirrors the computational power of Bitcoin miners (hash rate). Bitcoin miners certify ongoing transactions and uniqueness of the bitcoins by solving computationally demanding tasks and they obtain new (newly mined) bitcoins as a reward. Rewards and difficulties are given by a known formula.

Bitcoin is used mainly for two purposes -- purchases and exchange rate trading. Blockchain provides the total number of transactions and their volume excluding the exchange rate trading (exchange transactions). In addition, the ratio between volume of trade (mainly purchases) and exchange transactions is provided.

\subsubsection*{Exchanges}

Time series of exchange rates between BTC and various currencies are available at http://www.bitcoincharts.com. There, we obtain exchange volumes as a sum of four most important exchanges -- Bitfinex, Bitstamp, BTC-e and Mt. Gox -- which account for more than 90\% of all USD exchange transactions on the Bitcoin markets. Even though Mt. Gox is already in insolvency, we include it in the total exchange volume because it used to be the biggest exchange up till 2013 and its exclusion would strongly bias the actual volumes. After its bankruptcy, the volumes have converged to zero. For examination of the relationship between the USD and Chinese Renminbi (CNY) Bitcoin markets, we use prices and volumes of the btcnCNY market which is by far the biggest CNY exchange.

\subsubsection*{Search engines}

We utilize data provided by Google Trends at http://trends.google.com and by Wikipedia at http://http://stats.grok.se. For both, we are interested in term ``Bitcoin''. Google Trends standardly provide weekly data whereas Wikipedia series are daily. To obtain daily series for Google searches, one needs to download Google Trends data in three months blocks. The series are then chained and rescaled using the last overlapping month. 

\subsubsection*{Financial Stress Index}

The Financial Stress Index (FSI) is provided by the Federal Reserve Bank of Cleveland at https://www.clevelandfed.org/research/data/financial\_stress\_index/. FSI can be separated into various components. However, we use the overall index to control for all kinds of financial stress.

\subsubsection*{Gold price}

Gold prices for a troy ounce are obtained from https://www.gold.org/research and we use prices in Swiss francs (CHF) due to its stability and lack of expansive monetary policy. However, the results do not differ much regardless of the used currency.\\

According to Grinsted et al. \cite{Grinsted2004}, the series examined using the wavelet methodology should not be too far from the Gaussian distribution and mainly not multimodal. If the series in fact are multimode, it is suggested to transform them to a uniform distribution and in turn analyze quantiles of the original series. The inference based on the wavelet framework and related Monte Carlo simulations based significance is then reliable. For this matter, we transform all the original series accordingly as most of them, and majorly the Bitcoin prices, are multimodal and we thus interpret the results based on the quantile analysis. 

\subsection{Wavelets}

A wavelet $\psi_{u,s}(t)$ is a real-valued square integrable function defined as

\begin{equation}
\psi_{u,s}(t)=\frac{\psi\left(\frac{t-u}{s}\right)}{\sqrt{s}}
\end{equation}
with scale $s$ and location $u$ at time $t$. If the admissibility condition
\begin{equation}
C_{\Psi}=\int_0^{+\infty}{\frac{|\Psi(f)|^2}{f}df}<+\infty,
\end{equation}
where $\Psi(f)$ is the Fourier transform of a wavelet, holds, any time series can be reconstructed back from its wavelet transform. Wavelet has a zero mean and is standardly normalized so that $\int_{-\infty}^{+\infty}{\psi(t)}dt=0$ and $\int_{-\infty}^{+\infty}{\psi^2(t)}dt=1$. A continuous wavelet transform $W_x(u,s)$ is obtained via a projection of a wavelet $\psi(.)$ on the examined series $x(t)$ so that
\begin{equation}
W_x(u,s)=\int_{-\infty}^{+\infty}{\frac{x(t)\psi^{\ast}\left(\frac{t-u}{s}\right)dt}{\sqrt{s}}}
\end{equation}
where $\psi^{\ast}(.)$ is a complex conjugate of $\psi(.)$. The original series can be reconstructed from the continuous wavelet transforms for given frequencies so that there is no information loss \cite{Percival2000,Grinsted2004}. From a wide range of wavelets, we opt for the Morlet wavelet which provides a good balance between time and frequency localization \cite{Grinsted2004,Aguiar-Conraria2008}. 

%The Morlet wavelet with a central frequency $\omega_0$ is defined as
%\begin{equation}
%\psi(t)=\frac{e^{i\omega_0t-t^2/2}}{\pi^{1/4}}
%\end{equation}
%and for $\omega_0=6$, it provides a good balance between the time and frequency localization \cite{Grinsted2004}.

The continuous wavelet framework can be generalized for a bivariate case to study relationship between two series in time and across scales. A continuous wavelet transform is then generalized into cross wavelet transform as
\begin{equation}
W_{xy}(u,s)=W_x(u,s)W_y^{\ast}(u,s)
\end{equation}
where $W_x(u,s)$ and $W_y(u,s)$ are continuous wavelet transforms of series $x(t)$ and $y(t)$, respectively \cite{TorenceCompo98}. As the cross wavelet transform is in general complex, cross wavelet power $|W_{xy}(u,s)|$ is usually used as a measure of co-movement between the two series. Cross wavelet power uncovers regions in the time-frequency space where the series have common high power and it can be thus understood as a covariance localized in the time-frequency space. However, as for the standard covariance, the explanation power of $|W_{xy}(u,s)|$ is limited as it is not bounded.

To tackle this weakness, the wavelet coherence is introduced as

\begin{equation}
R_{xy}^2(u,s)=\frac{|S\left(\frac{1}{s}W_{xy}(u,s)\right)|^2}{S\left(\frac{1}{s}|W_x(u,s)|^2\right)S\left(\frac{1}{s}|W_x(y,s)|^2\right)},
\end{equation}
where $S$ is a smoothing operator \cite{Grinsted2004,TorenceWebster97}. The squared wavelet coherence ranges between 0 and 1 and it can be interpreted as a squared correlation localized in time and frequency. Due to the above mentioned complexness of the used wavelets and in turn the use of the squared coherence rather than coherence itself, the information about a direction of the relationship is lost. For this purpose, a phase difference is introduced as
\begin{equation}
\varphi_{xy}(u,s)=\tan^{-1}\left(\frac{\mathfrak{I}\left[S(\frac{1}{s}W_{xy}(u,s))\right]}{\mathfrak{R}\left[S(\frac{1}{s}W_{xy}(u,s))\right]}\right),
\end{equation}
where $\mathfrak{I}$ and $\mathfrak{R}$ represent an imaginary and a real part operator, respectively. Graphically, the phase difference is represented by an arrow. If the arrow points to the right (left), the series are positively (negatively) correlated, i.e. they are in the in-phase or the anti-phase, respectively, and if the arrow points down (up), the first series leads the other by $\frac{\pi}{2}$ (vice versa). The relationship is usually a combination of the two, i.e. if the arrow points to north-east, the series are positively correlated and the second series leads the first one. Note that the interpretation of phase relationships is partially dependent on specific expectations about the relationship because a leading relationship in the in-phase can easily be a lagging relationship in the anti-phase. Please refer to Ref. \cite{Grinsted2004} for a detailed description.

Recently, the partial wavelet coherence has been proposed to control for common effects of two variables on the third one \cite{Mihanovic2009,Ng2012} and it is defined as
\begin{equation}
RP_{y,x_1,x_2}^2=\frac{|R_{yx_1}-R_{yx_2}R_{yx_1}^{\ast}|^2}{\left(1-R_{yx_2}^2\right)^2\left(1-R_{x_2x_1}^2\right)^2}.
\end{equation}
The partial wavelet coherence ranges between 0 and 1 and it can be understood as the squared partial correlation between series $y(t)$ and $x_1(t)$ after controlling for the effect of $x_2(t)$ localized in time and frequency.

\section{Results}

We analyze drivers of the exchange rate between Bitcoin (BTC) and the US dollar (USD) between 14.9.2011 and 28.2.2014. This specific exchange rate pair is selected because trading volumes on the USD markets form a strong majority followed with a profound lag by the Chinese renminbi (CNY). The analyzed period is restricted due to availability of Bitcoin price index covering the most important USD exchanges. Note that an analysis of a specific exchange is not feasible as historically the most important market, Mt. Gox, was filed for bankruptcy after serious problems with bitcoins withdrawals in 2014. For this reason, we use the CoinDesk Bitcoin price index (BPI) which is constructed as an average price of the most liquid exchanges. Please refer to the Methods section for more details about BPI. 

\begin{figure}[htbp]
\center
\begin{tabular}{c}
\includegraphics[width=6.5in]{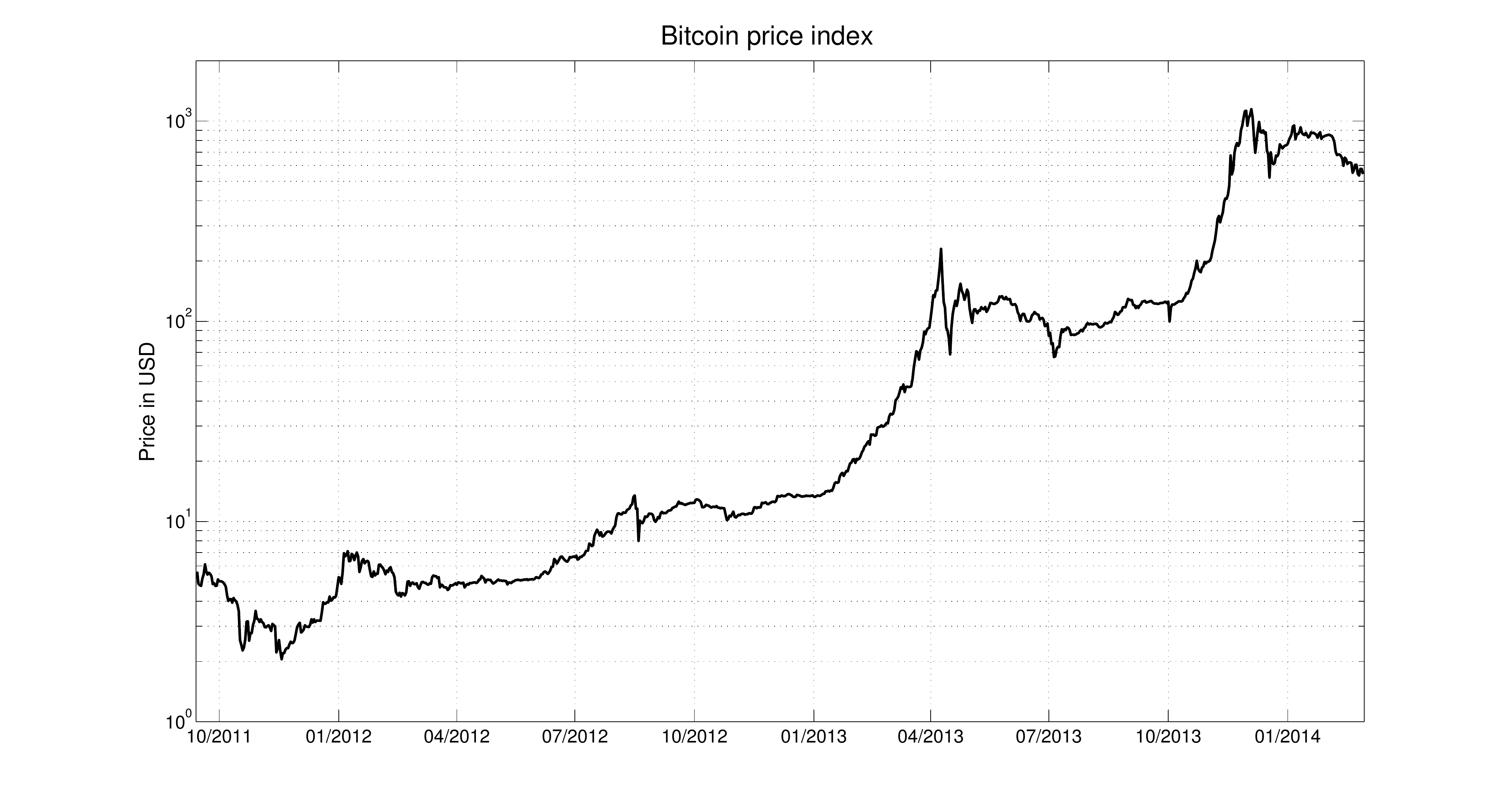}\\
\end{tabular}
\caption{\footnotesize\textbf{Bitcoin price index.} Values of the index are shown in the USD (for the USD markets) and in the logarithmic scale.\label{fig1}}
\end{figure}

Evolution of the price index is shown in Fig. \ref{fig1} where we observe that the Bitcoin price is dominated by episodes of explosive bubbles followed by corrections which, however, never return back to the starting values of the pre-bubble phase. The analyzed period starts at the value of approximately \$5 per a bitcoin and ends at approximately \$600. Even though the most recent dynamics of the Bitcoin price can be described as a slow decreasing trend, a potential profit of a buy-and-hold strategy of almost 12000\% in less than 30 months remains appealing.

Compared to the standard currencies such as the US dollar, the Euro, the Japanese Yen, and others, Bitcoin shines due to an unprecedented data availability. It is completely unrealistic to know the total amount of the US dollars in the worldwide economy on a daily basis. In a similar manner, it is even impossible to track the number of transactions taking place with a use of USD and others. However, Bitcoin provides such information on daily basis, publicly and freely. Such data availability allows for more precise statistical analysis. We look at the Bitcoin prices from various aspects that might influence the price or that are usually gossiped about as drivers of the Bitcoin exchange rate. We start with the economic drivers, or potential fundamental influences, followed by transaction and technical drivers, influences of interest in Bitcoin, its possible safe haven status and finally, we focus on the effects of the Chinese Bitcoin market.

%%%%%%%%%%%

%\subsection{Fundamentals}
\subsection{Economic drivers}

In economic theory, price of a currency is standardly driven by its use in transactions, by its supply and by the price level. Time series for all these variables are available or we are able to reconstruct them from other series, see the Methodology section for more details. 

\begin{figure}[htbp]
\center
\begin{tabular}{c}
\includegraphics[width=6.5in]{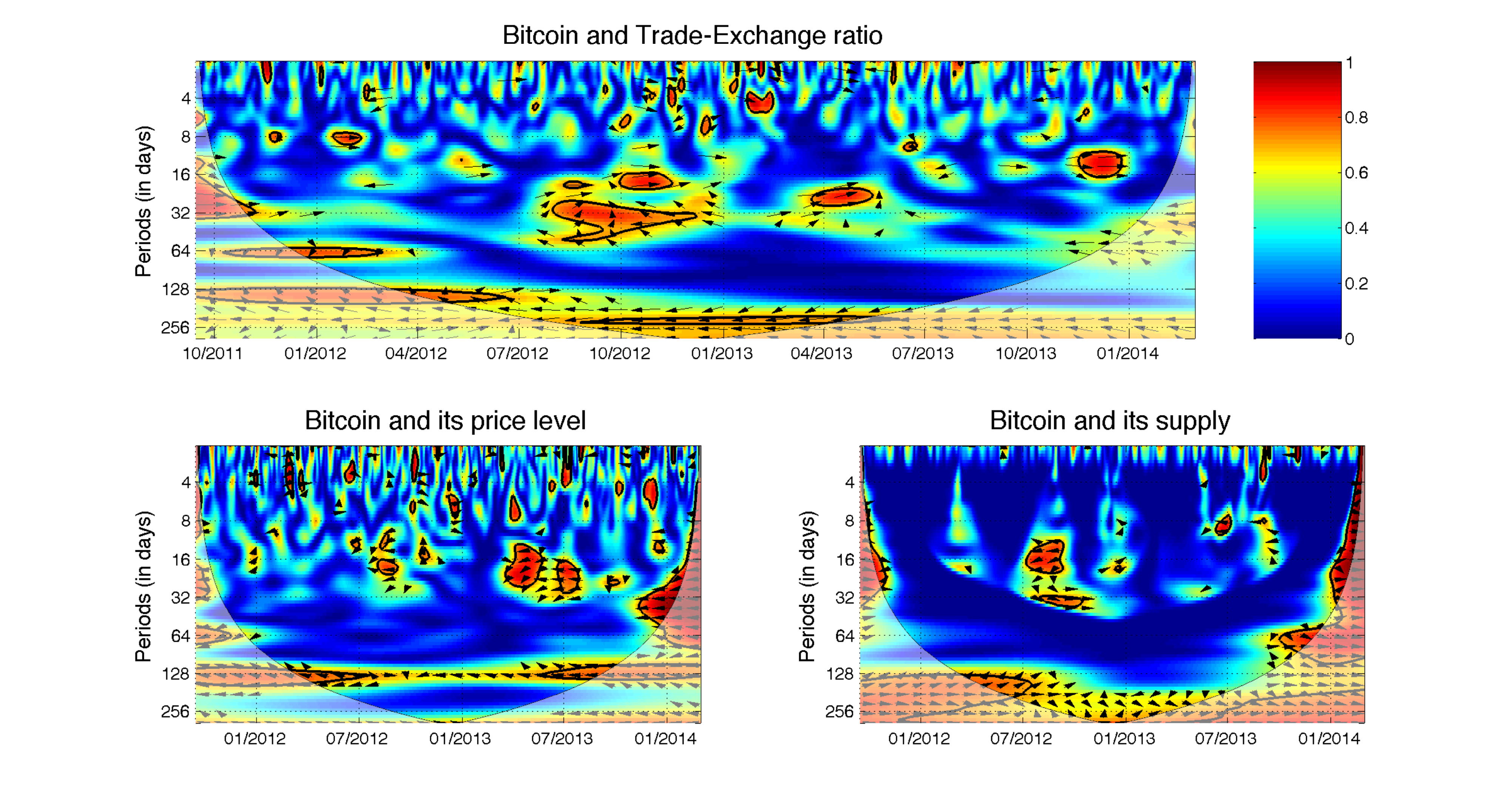}\\
\end{tabular}
\caption{\footnotesize\textbf{Fundamental drivers.} Wavelet coherence is represented by colored contour for which it holds that the hotter the color the higher the local correlation in the time-frequency space (with time on the $x$-axis and scale on the $y$-axis). Matching of colors and correlation levels are represented by the scale on the right hand side of the upper graph. Regions with significant correlations tested against the red noise are contrasted by a thick black curve. Cone of influence separating the regions with reliable and less reliable estimates is represented by bright and pale colors, respectively. Phase (lag-lead) relationships are shown by the arrows -- positive correlation is represented by an arrow pointing to the right, negative correlation with one to the left, leadership of the first variable is shown by a downwards pointing arrow and if it lags, the relationship is represented by an upward pointing arrow. The last two relationships hold for the in-phase relationship (positive correlation) and for the anti-phase (negative correlation), it holds vice versa. Now specifically for the fundamental drivers. Bitcoin price is negatively correlated to the Trade-Exchange ratio \textit{(top)} in the long-term for the whole analyzed period and there is no evident leader in the relationship. The Bitcoin price level is negatively correlation with the Bitcoin price in the long-term for the whole analyzed period as well \textit{(bottom left)} with no evident leader. Between relatively calm period between 05/2013-09/2013, the price level leads the prices in the medium term. Supply of bitcoins is positively correlated with the price in the long-term \textit{(bottom right)} with no evident leader. \label{fig2}}
\end{figure}

As a measure of the transactions use, i.e. demand for the currency, we use the ratio between trade and exchange transactions volume, which we abbreviate to Trade-Exchange ratio. The ratio thus shows what is the ratio between volumes on the currency exchange markets and in trade (purchases, services, etc.). Therefore, the lower the ratio the more frequently Bitcoin is used for ``real world'' transactions. From the theory, the price of the currency should be positively correlated with its usage for real transactions, as it increases the utility of holding the currency, and the usage should be leading the price. In Fig. \ref{fig2}, we show the squared wavelet coherence between the Bitcoin price and the ratio. We thus see the evolution of the local correlation in time and across frequencies. The hotter the color the higher the correlation. Statistically significant correlations are highlighted by a thick black curve around the significant regions, significance is based on the Monte Carlo simulations agains the null hypothesis of the red noise, i.e. autoregressive process of order one. Cone of influence separates the reliable (full colors) and less reliable (pale colors) regions. Phase difference, i.e. a lag or lead relationship is represented by oriented arrows. Please refer to the Methods section for more details. Specifically for the Trade-Exchange ratio, we observe strong yet not statistically significant at 5\% level relationship at high scales. The variables are in the anti-phase so that they are negatively correlated in the long term. However, there is no strong leader in the relationship. Slightly dominating frequency of the arrows pointing to the south-west hints that the ratio is a weak leader. On the shorter scales, most of the arrows point to the north-east indicating that the variables are positively correlated and the prices lead the Trade-Exchange ratio. Note that this relationship is visible mainly for the periods of extreme price increases of BTC. In words, Bitcoin appreciates in the long run if it is used more for trade, i.e. non-exchange, transactions, and the increasing price boosts the exchange transactions in the short run. The former is thus in hand with the theoretical expectations and the latter shows that increasing prices -- potential bubbles -- boost demand for the currency at the exchanges. 

Price level is an important factor due to an expectation that goods and services should be available for the same, or at least similar, prices everywhere and the misbalances are controlled for by the exchange rate. When price level associated with one currency decreases with respect to price level of the other currency, the first currency should be appreciating and its exchange rate should be thus increasing. An expected causality goes from the price level to the exchange rate (price) of Bitcoin. Price level in our case is constructed as an average price of a trade transaction for a given day. Fig. \ref{fig2} uncovers that the most stable interactions take place at high scales around 128 days. The relationship is negative as expected but the leader is not clear. There is also a significant region at lower scales around one month between 04/2013 and 07/2013. The relationship is again negative as expected but a leadership of the price level is more evident here. Most of the other significant correlations are outside of the reliable region. 

Money supply works as a standard supply so that its increase leads to price decrease. Negative relationship is thus expected. Moreover, due to a known algorithm of bitcoins creation, only long-term horizons are expected to play a role. In Fig. \ref{fig2}, we observe that there is some relationship between Bitcoin price and its supply. However, most of the significant regions are outside of the reliable region. Moreover, the orientation of the phase arrows is unstable so that it is not possible to detect either of a sign or a leader of the relationship. This might be due to the fact that both current and future money supply is known in advance so that its dynamics can be easily included in the expectations of Bitcoin users and investors.

\subsection{Transaction drivers}

\begin{figure}[htbp]
\center
\begin{tabular}{c}
\includegraphics[width=6.5in]{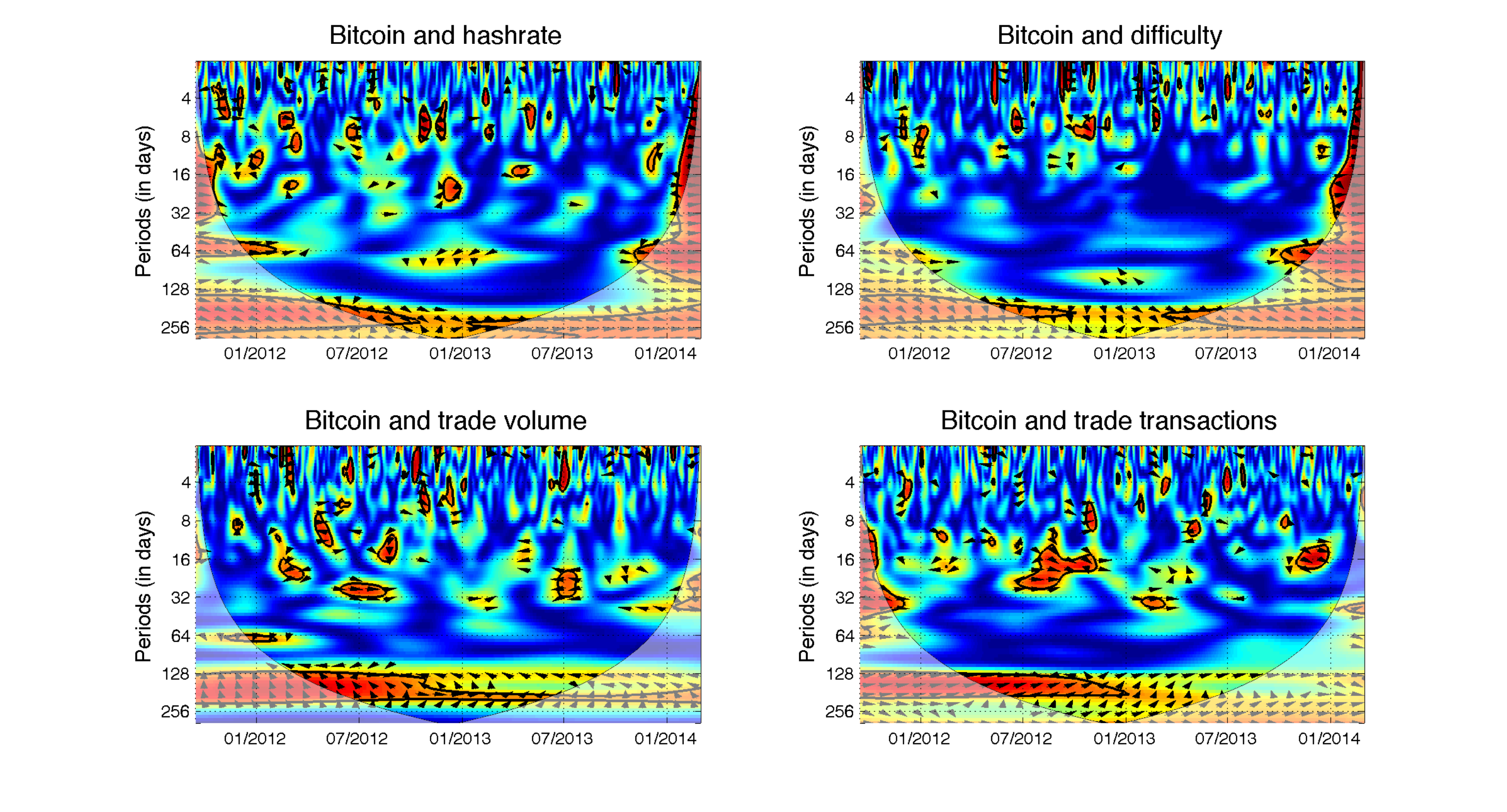}\\
\end{tabular}
\caption{\footnotesize\textbf{Currency mining and trade usage.} Description and interpretation of relationships hold from Fig. \ref{fig2}. Both hash rate \textit{(top left)} and difficulty \textit{(top right)} are positively correlated with the Bitcoin price in the long-term. The price leads both relationships as the phase arrow points to south-east in most cases and the interconnection remains quite stable in time. The trade volume \textit{(bottom left)} is again connected to the Bitcoin price mainly in the long-term. However, the relationship is not much stable in time time. Up to 10/2012, we observe negative correlation between the two and the price is the leader. The relationship then becomes less significant and the leader position is not evident anymore.For the trade transactions \textit{(bottom right)}, the relationship is positive in the long-term and the transactions are leading the Bitcoin prices. However, the relationship becomes weaker in time and it is not statistically significant from 01/2013. \label{fig3}}
\end{figure}

Usage of bitcoins in real transactions is tightly connected to the fundamental aspects of its value. However, there are possibly two contradictory effects between usage of bitcoins and their price which might be caused by its speculative aspect. One effect stems in a standard expectation that the more the coins are used the higher the demand for them and thus the higher the price. However, if the prices are driven by speculations, their volatility and uncertainty about the price as well as an increasing USD value of transaction fees can lead to negative relationship. As measures of usage, trade volume and trade transactions are used. In Fig. \ref{fig3}, we observe that for both variables, the significant relationships take place mainly at higher scales and also mainly in 2012. The effect diminishes in 2013 and at lower scales, the significant regions are only short-lived and can be due to statistical fluctuations and noise. For the trade transactions, it is clear that the relationship is positive and that the transactions lead the prices, i.e. increasing usage of Bitcoin in real transactions leads to an appreciation of Bitcoin in long run. However, the effect becomes weaker in time. For the trade volume, the relationship changes in time and the phase arrows change their direction too often to give us any strong conclusion.

\subsection{Technical drivers}

Bitcoins are mined according to a given algorithm so that the planned supply of bitcoins is kept. Miners, who mine new bitcoins as a reward for certification of transactions in blocks, thus provide an inflow of new bitcoins into circulation. However, mining is contingent to solving computationally very demanding problem. Moreover, to keep the creation of new bitcoins in check and according to the planned formula, the difficulty of solving the rewarding problem increases according to a computational power of current miners. The difficulty is then given by minimal needed computational efficiency of miners and it reflects current computational power of the system measured in hashes. Hash rate then becomes another measure of system productivity which is reflected in the system difficulty which is recalculated every 2016 blocks of 10 minutes, i.e. approximately two weeks. This way the Bitcoin supply remains balanced and the system is not flooded with bitcoins. The Bitcoin mining is thus an investment opportunity where the computational power is changed for bitcoins. The mining itself is connected with costs of an investment into hardware as well as electricity. Note that possibility of the Bitcoin mining (and also mining of other mining-based crypto-currencies) led to development and production of hardware specifically designed for the coins mining. This has led to an increasing costs of mining, soaring mining hash rate and difficulty which has gradually drifted small miners away from the pool. 

There are again two opposing effects between the Bitcoin price and the mining difficulty as well as the hash rate. Mining can be seen as a kind of investment into Bitcoin. Rather than buying bitcoins directly, investor invests into the hardware and obtains the coins indirectly. This leads to two possible effects. Increasing price of Bitcoin can motivate market participants to start investing into hardware and start mining, which leads to increased hash rate and in effect also to a higher difficulty. Or, increasing hash rate and difficulty connected with increasing cost demands for hardware and electricity drive more miners off the mining pool. If these miners used to mine the coins as an alternative to the direct investment, they can transfer to Bitcoin purchasers and thus increase demand for bitcoins and in turn their price as well.

Fig. \ref{fig3} summarizes the wavelet coherence for both hash rate and difficulty. We observe very similar results for both measures as expected as these two are very tightly interwoven. Both measures of the mining difficulty are positively correlated with the price at high scales, i.e. in the long run. The relationship is clearer for the difficulty which shows that the Bitcoin price leads the difficulty even though the leadership becomes weaker in time. The effect of rising prices attracting new miners thus seems to dominate the relationship. The weakening of the relationship in time can be attributed to the current stable or slowly decreasing price of bitcoins which no longer offsets the costs of computational power needed for successful mining in a sufficient manner.

\subsection{Interest}

One of possible drivers of the Bitcoin prices is its popularity. Simply put -- increasing interest in the currency, connected with a simple way of actually investing into it, leads to an increasing demand and thus increasing prices. To quantify the interest in Bitcoin, we utilize Google and Wikipedia engines search queries for word ``Bitcoin'". It is obviously hard to distinguish between various incentives of internet users searching for information about Bitcoin. Nonetheless, we assume that an increased interest leads to increasing prices.

\begin{figure}[htbp]
\center
\begin{tabular}{c}
\includegraphics[width=6.5in]{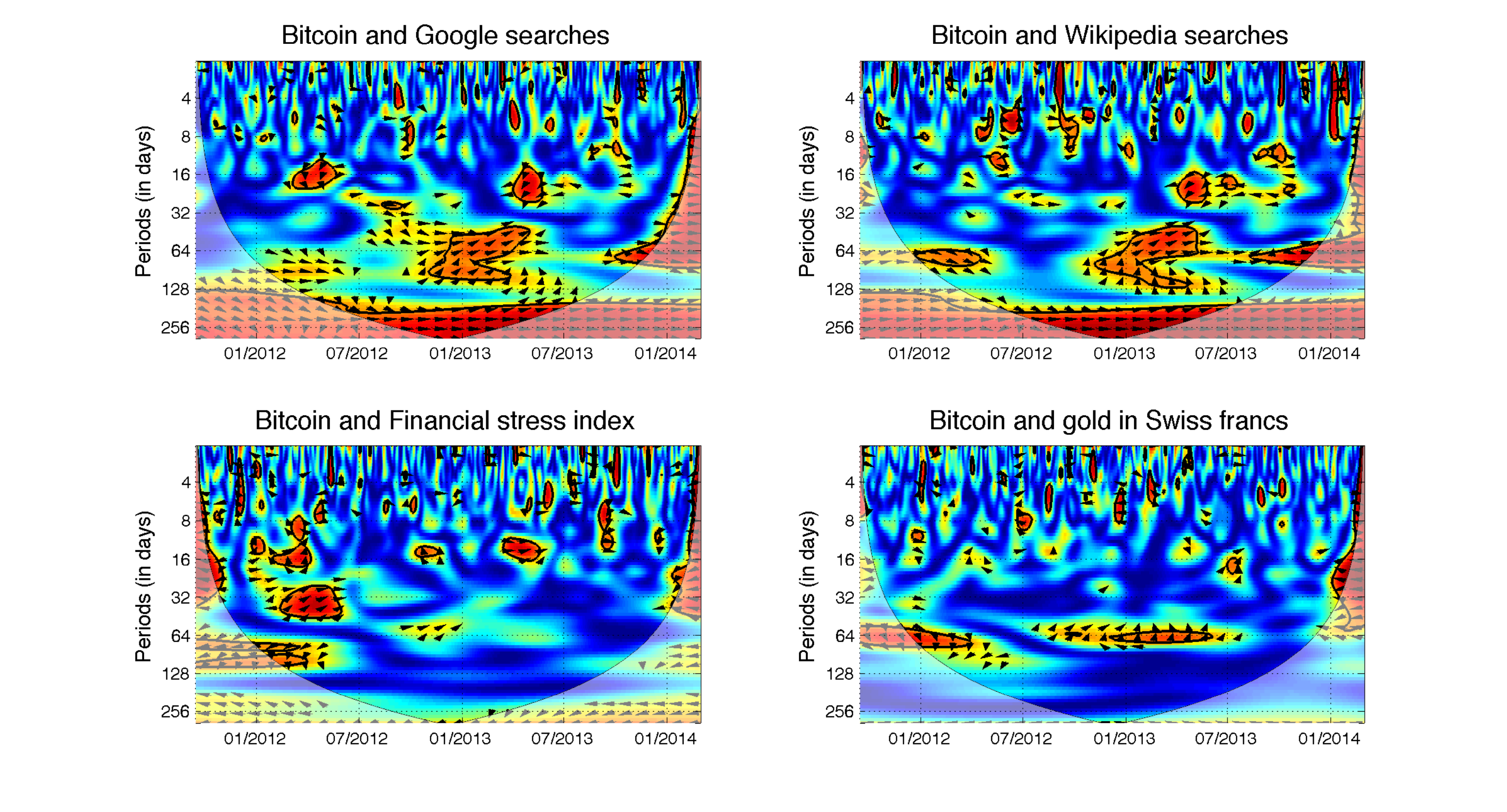}\\
\end{tabular}
\caption{\footnotesize\textbf{Search engines and safe haven value.} Description and interpretation of relationships hold from Fig. \ref{fig2}. Searches on both engines \textit{(top)} are positively correlated with the Bitcoin price in the long run. For both, we observe that the relationship somewhat changes in time. In the first third of the analyzed period, the relationship is lead by the prices whereas in the last third of the period, the search queries lead the prices. Unfortunately, the most interesting dynamics remains hidden in the cone of influence and it is thus not too reliable. Apart from the long run, there are several significant episodes at the lower scales with varying phase direction hinting that the relationship between search queries and prices depends on the price behavior. Moving to the safe haven region, we find no strong and lasting relationship between the Bitcoin price and either the Financial stress index \textit{(bottom left)} or gold price \textit{(bottom right)}. The significant regions at medium scales for gold are rather connected to the dynamics of the Swiss franc exchange rate.\label{fig4}}
\end{figure}

In Fig. \ref{fig4}, we show the wavelet coherence between Bitcoin prices and search engines queries. We observe that both engines give very similar information. The co-movement is most dominant at high scales. However, we observe that the relationship changes in time. Up to the half of 2012, the prices lead the interest and such relationship is more evident for the Google searches. The directionality of the relationship becomes weaker and starting from the beginning of 2013, it is hard to tell the leader confidently even though the searches tend to boost the prices. Nonetheless, the leadership is not very apparent. Apart from the long-term relationship, there are other interesting periods when the interest in the coins and the prices are interconnected. The most visible of these takes place between 01/2013 and 04/2013 at medium scales between approximately 30 and 100 days. The prices are evidently led by the interest in Bitcoin during this period. Note that the first quarter of 2013 was connected to an exploding bubble when Bitcoin rocketed from \$13 to above \$200. Similar dynamics seems to be present also for the other bubble starting in 10/2013. Unfortunately, the whole development of this latter bubble is hidden in the cone of influence and the findings are thus not statistically reliable. Back to the 01/2013 - 04/2013 bubble, its deflation is also connected to an increased interest of the internet users. The interest and prices are then negatively correlated and the interest still leads the relationship. However, the correlations are found at lower scales than for the bubble formation. The interest in Bitcoin thus seems to have an asymmetric effect during the bubble formation and its bursting -- during the bubble formation, the interest boosts the prices further, and during the bursting, it pushes them lower. Moreover, the interest influence happens at different frequencies during the bubble formation and its bursting so that the increased interest has a more rapid effect during the price contraction than during the bubble build-up.

\subsection{Safe haven}

Even though it might seem as an amusing notion, Bitcoin has been also labeled as a safe haven investment at one point. This point was the Cypriot economic and financial crisis taking place in the beginning of 2012. There were speculations that some of the funds from the local banks were transferred to the Bitcoin accounts assuring their anonymity. Leaving these speculations aside, we analyze the possibility of Bitcoin being the safe haven quantitatively. Specifically, we examine the relationship of the Bitcoin prices with the Financial Stress Index (FSI) and gold price in the Swiss francs. The former is a general index of financial uncertainty. The latter is chosen due to the fact that gold is usually labelled as a long-term storage of value and the Swiss franc is considered as a very stable currency, being frequently labeled as the safe haven itself. If Bitcoin is really the safe haven, it would be positively correlated with both utilized series.

Fig. \ref{fig4} summarizes the results. For the FSI, we observe that there is actually only one period of time which shows interesting interconnection between the index and the Bitcoin prices. The period is exactly the one of the Cypriot crisis and most of the co-movements are observed at scales around 30 days. Increasing FSI leads the Bitcoin prices up. However, apart from the Cypriot crisis, there are no longer-term time intervals where the correlations are both statistically significant and reliable (in a sense of the cone of influence). Turning now to the gold price, there seems to be practically no relationship apart from two significant islands at scales around 60 days. However, these are most probably connected to the dynamics of  gold itself because the first significant period coincides with a rapid increase of gold price culminating around September 2011 (big proportion of the significant region is outside of the reliable part of the coherence) and the second collides with stable decline of gold prices. It thus seems that Bitcoin is not much connected to the dynamics of gold but even more, it is not obvious whether gold still remains the safe haven as it is believed to be. Either way, we find no signs of Bitcoin being a safe haven.

\subsection{Influence of China}

There are claims that events happening on the Chinese Bitcoin market have a significant impact on the USD markets. Truly, some of the extreme drops as well as price increases in the Bitcoin exchange rate coincide with dramatic events in China and Chinese regulations concerning Bitcoin. Probably the most notable example is the development around Baidu, which is an important player in the Chinese online shopping. Announcement that Baidu is accepting bitcoins in mid-October 2013 started a surge in its value which was, however, cut back by Chinese regulation banning bitcoins from electronic purchases in early-December 2013. The Chinese market is thus believed to be an important player in the digital currencies and especially in Bitcoin. To examine the relationship between the Chinese renminbi (CNY) and the US dollar markets, we look at their prices and exchange volumes.

\begin{figure}[htbp]
\center
\begin{tabular}{c}
\includegraphics[width=6.5in]{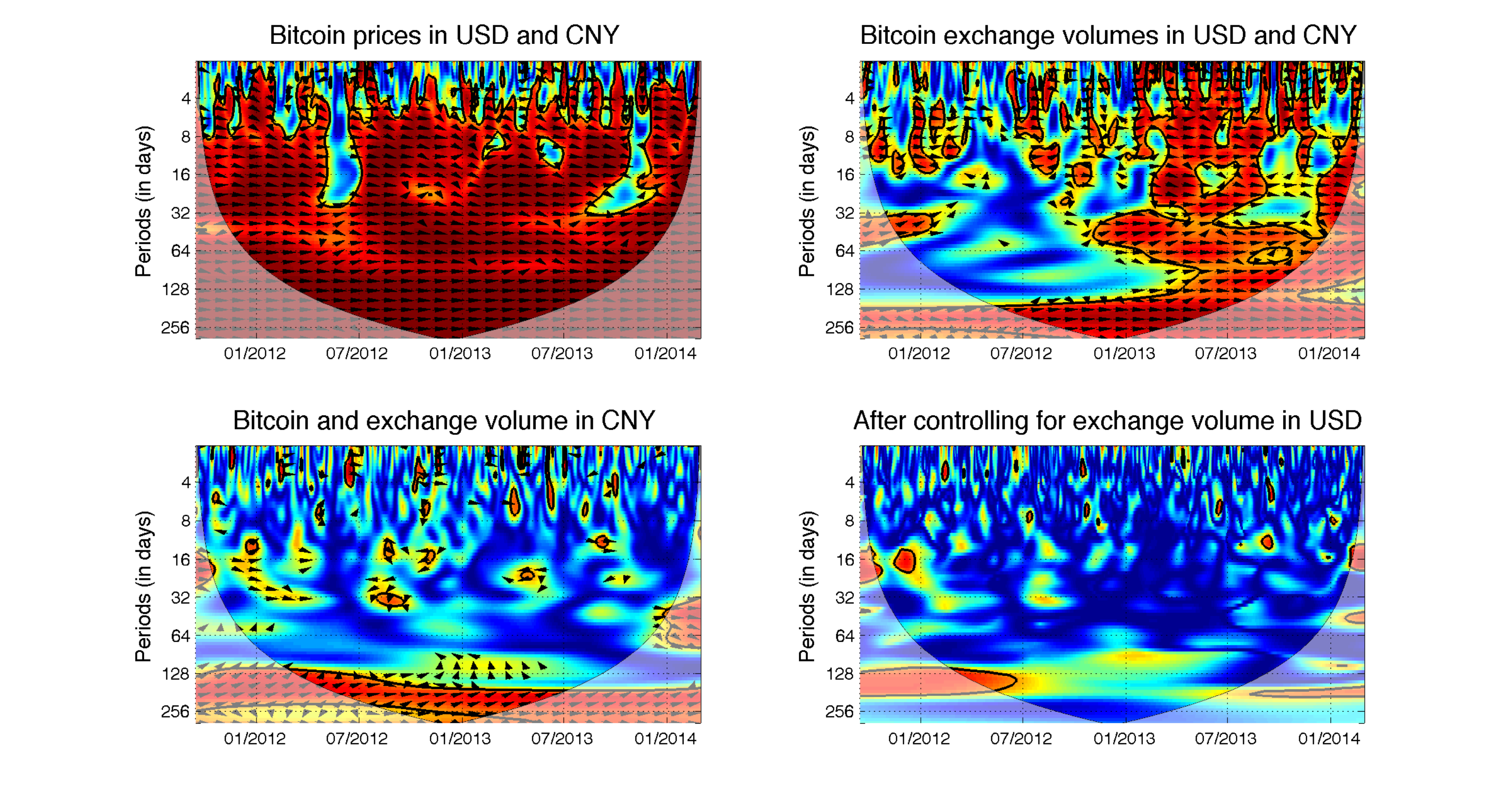}\\
\end{tabular}
\caption{\footnotesize\textbf{Influence of Chinese market.} Description and interpretation of relationships hold from Fig. \ref{fig2}. Bitcoin prices in USD and CNY \textit{(top left)} move together at almost all scales and during the whole examined period. There is no evident leader in the relationship, even though the USD market seems to slightly lead the CNY one at lower scales. However, at the lowest scales (the highest frequencies), the correlations vanish. For the exchange volumes \textit{(top right)}, the two markets are strongly positively correlated at high scales. However, for the lower scales, the correlations are significant only from the beginning of 2013 onwards. There is again no dominant leader in the relationship. The CNY exchange volume then leads the USD prices in the long run \textit{(bottom left)}. However, when we control for the effect of the USD exchange volume \textit{(top right)}, we observe that the correlations vanish.\label{fig5}}
\end{figure}

Fig. \ref{fig5} covers all interesting results. Prices of both markets are tightly connected and we observe strong positive correlations at practically all scales and during the whole examined period. From the phase arrows, we can hardly find a leader in the relationship. More interesting dynamics is found for the exchange volumes. Here, we find that the volumes are strongly positively correlated as well but only from the beginning of 2013 onwards. Before that period, the interconnections are visible only at the highest scales and most of the dynamics falls outside of the reliable region. Note that the trading volumes on the CNY market were quite low during 2012. In the significant part, we again find that the relationship is strong and it is not easy to find an evident leader. Nonetheless, the period between 10/2013 and 12/2013 is again connected to decoupling of the markets in the same way as for the prices. From these results, we can conclude that both markets tend to move together very tightly both in prices and in volumes.

One might believe that if the Chinese market is an important driver of the BTC exchange rate with USD, an increased exchange volume in China might increase demand in all markets so that the Chinese volume and the USA price would be connected. This is even more stressed by the fact that shorting (selling now and buying later) of bitcoins is still limited. In Fig. \ref{fig5}, we show that this is indeed true and the relationship is present at the high scales again. As most of the phase arrows point to the north-east region, the Chinese volume leads the USD prices. However, as we have discussed above, the USD and CNY exchange volumes are strongly correlated, and at the high scales, this is true for the whole analyzed period. Therefore, the relationship between CNY volume and USD price might be spuriously found due to such correlation. To control for this effect, we utilize the partial wavelet coherence which filters this effect away. In the last chart of Fig. \ref{fig5}, we show that after controlling for the exchange volume of the USD market, practically no interconnection between the CNY volume and the USD price remains. All in all, we find no causal relationship between the CNY and the USD markets in the analyzed dataset.

\section{Discussion}

Bitcoin price dynamics has been a controversial topic since the crypto-currency became more popular and known to a wider audience. We have tackled the issue of the Bitcoin price formation and development from a wider perspective and we have investigated the most frequently claimed drivers of the prices. There are several interesting findings. First, even though Bitcoin is usually labelled as a purely speculative asset, we find that standard fundamental factors -- usage in trade, money supply and price level -- play a role in Bitcoin prices in the long term. Second, from the technical standpoint, the increasing prices of Bitcoin motivate users to become miners. However, the effect is found to be vanishing in time as the specialized mining hardware components have driven the hash rates and difficulty too high. Nonetheless, this is a standard market reaction to an obvious profit opportunity. Third, the prices of bitcoins are driven by investors' interest in the crypto-currency. The relationship is most evident in the long run but during the episodes of explosive prices, the interest drives the prices further up, and during rapid declines, it pushes them further down. Fourth, Bitcoin does not seem to be a safe haven investment. And fifth, even though the USD and CNY markets are tightly connected, we find no clear evidence that the Chinese market influences the USD market.

%\section*{Methods}

\section*{Acknowledgements}
The research leading to these results has received funding from the European Union's Seventh Framework Programme (FP7/2007-2013) under grant agreement No. FP7-SSH-612955 (FinMaP) and the Czech Science Foundation project No. P402/12/G097 ``DYME -- Dynamic Models in Economics''. Thanks also go to Aslak Grinsted and Eric Ng for providing the MatLab wavelet packages.

%\newpage

\bibliographystyle{vancouver}
%\bibliography{BitCoin}

%\section*{Data accessibility}

%\section*{Author contributions}
%L.K. solely wrote the main manuscript text, prepared the figures and reviewed the manuscript.

%\section*{Additional information}
%\textbf{Competing financial interests:} The author declares no competing financial interests.\\
%\textbf{Data retrieval:} CoinDesk BPI series was retrieved on 17 March 2014. Time series from http://www.blockchain.info were retrieved on 22 March 2014. And time series from http://www.bitcoincharts were retrieved on 24 March 2014.
%Search volume data were retrieved by accessing the Google Trends website (http://www.google.com/trends) on 5 July 2013 and the Wikipedia article traffic statistics site (http://stats.grok.se) on 21 August 2013. \textit{BitCoin} series were obtained from http://www.bitcoincharts.com between 5.-8.7.2013.

\newpage

\end{document}